\documentclass[epj]{svjour}

\usepackage{graphicx}
\usepackage{fancyhdr}
\usepackage{amsmath,amssymb}
\def\makeheadbox{{
 }}
\def\mytitle{My title} 
\def\myauthors{My name}  
\def\mytype{My type of session}
\def\mysession{My session}

\def\mytitle{Little Higgs Model Discrimination at the LHC and ILC}
\def\myauthors{J\"urgen Reuter}    
\def\mytype{Parallel Talk}    
\def\mysession{Alternatives}

\begin{document}
\title{Little Higgs Model Discrimination at the LHC and ILC}
\author{J\"urgen Reuter
\thanks{\emph{Email:} juergen.reuter@desy.de}%
}
\institute{University of Freiburg, Institute
 of Physics, Germany
}
%


\date{}
\abstract{
We propose a means to discriminate between the two basic variants of
Little Higgs models, the Product Group and Simple Group models, at the
next generation of colliders. It relies on a special coupling of light
pseudoscalar particles present in Little Higgs models, the
pseudoaxions, to the $Z$ and the Higgs boson, which is present only in
Simple Group models. We discuss the collider phenomenology of the
pseudoaxion in the presence of such a coupling at the LHC, where
resonant production and decay of either the Higgs or the pseudoaxion
induced by that coupling can be observed for much of parameter
space. The full allowed range of parameters, including regions where
the observability is limited at the LHC, is covered by a future ILC,
where double scalar production would be a golden channel to look
for. 
} 
\maketitle

\section{Pseudoaxions in Little Higgs Models}

Little Higgs Models~\cite{lhm_moose} provide a
solution to the hierarchy problem, as they stabilize the Higgs boson
against quadratic divergences at the one-loop level by the mechanism
of collective symmetry breaking: the Higgs is charged under two global 
symmetry groups, which both need to be broken in order to lift the
flat direction in the potential of the Higgs boson and make it a
pseudo-Nambu-Goldstone boson (PNGB). 
Collective breaking models can be classified in three different
categories, the so-called moose models with a moose diagram structure
of links of global and local symmetry groups, the product-group models
and the simple-group models. In the product-group models (the
most-studied case is the Littlest Higgs) the electroweak gauge group
is doubled, broken down to the group $SU(2)_L$, while the Higgs shares
together with the other PNGBs an irreducible representation of the
coset space of the symmetry breaking. On the other hand, in
simple-group models the electroweak gauge group is enlarged to a
simple $SU(N)$ group, while the Higgs is distributed over several
multiplets of the global symmetry group, which usually has a product
group structure similar to chiral symmetries in QCD~\cite{simple}. For
an overview see~\cite{review}. 

\begin{figure}
\centerline{\includegraphics[width=0.8\columnwidth]
  {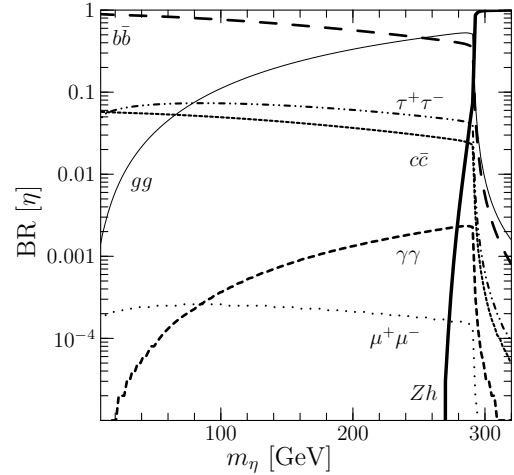}}   
\caption{Branching ratios of the pseudoaxion $\eta$ in the Simplest
  Little Higgs as a function of its mass.}\label{fig:etabr}  
\end{figure} 

The two crucial scales in the Little Higgs set-up are the cut-off
scale $\Lambda$ where the models are embedded in a UV-complete theory
(usually a strongly-interacting theory with a partonic substructure of
the PNGBs) and the intermediate scale $F$ which determines the masses
and decay constants of the PNGBs (except for the Higgs which is down
at $v$ by the collective symmetry breaking mechanism). Electroweak
precision observables and direct search limits~\cite{lowenergy} tell
us that the scale $F$ must be at least of the order of $1-2$
TeV. Paradoxically, the Higgs boson in Little Higgs models tends to be
quite heavy compared to the Standard Model or the MSSM, of the order
of $200-600$ GeV~\cite{little_kr}. For Little Higgs model scales that
high most new particles will be produced close to the kinematical limit at
the LHC, such that a precision determination of their parameters might
be difficult. Furthermore, also the sensitivity of the ILC in indirect
measurements might be limited, if the new physics does couple to SM
fermions only very weakly~\cite{resonances}. A method to distinguish
between different models, especially at the LHC, is highly
welcome. Such a method will be presented here.   

Little Higgs models generally have a huge global symmetry group, which
contains not only products of simple groups but also a certain number
of $U(1)$ factors. These Abelian groups can either be gauged, in which
case they lead to a $Z'$ boson, or they are only (approximate) global
symmetries. In the latter case there is a PNGB
attached to that spontaneously broken global $U(1)$
factor~\cite{pseudoaxions}. The number of pseudoaxions in a given
model is determined by the mismatch between the rank reduction in
the global and the local symmetry group, since it gives the number of
uneaten bosons. In the Littlest Higgs, e.g., there is one such
pseudoaxion, in the Simplest Little Higgs~\cite{simplest} there is
one, in the original simple group model there are two, in the minimal
moose model there are four, and so on.

These particles are electroweak singlets, hence all couplings to SM
particles are suppressed by the ratio of the electroweak over the
Little Higgs scale, $v/F$. There mass lies in the range from several
GeV to a few hundred GeV, being limited by a naturalness argument and
the stability of the Coleman-Weinberg potential. For the Simplest
Little Higgs, on whose phenomenology we will concentrate here, there
is a seesaw between the Higgs and the pseudoscalar
mass~\cite{pseudoaxions}, determined by the explicit symmetry breaking
parameter $\mu$, where $m_\eta \approx \sqrt{2} \mu$.  Since the
pseudoaxions $\eta$ inherit the Yukawa coupling
structure from the Higgs bosons, they decay predominantly to the
heaviest available fermions in the SM, and because of the absence of
the $WW$ and $ZZ$ modes, the anomaly-induced decays $gg$ and
$\gamma\gamma$ are sizable over a wide mass range,
cf.~Fig.~\ref{fig:etabr}. From this, one can see that as soon as the
decay to $HZ$ is kinematically allowed, it dominates completely. Such
a $\eta HZ$ coupling, which is possible only after electroweak
symmetry breaking and hence proportional to $v/F$, is only allowed in
simple group models and is forbidden to all orders in product group
models. One can factor out the $U(1)_\eta$ group from the matrix of
PNGBs. We use $\xi=\exp\left[i\eta/F\right]$ for the
pseudoaxion field and $\Sigma=\exp\left[i\Pi/F\right]$ for the
non-linear representation of the remaining Goldstone multiplet $\Pi$
of Higgs and other heavy scalars.  Then, for product group models, the
kinetic term may be expanded as
\begin{align}
  \label{LL-kin}
  \mathcal{L}_{\text{kin.}} &\sim\; F^2
  \text{Tr} \left[ (D^\mu (\xi \Sigma)^\dagger (D_\mu (\xi \Sigma)) \right]
  \nonumber \\
  &=\; \ldots - 2F(\partial_\mu \eta)\,
  \text{Im} \text{Tr} \left[  (D^\mu \Sigma)^\dagger \Sigma \right]
  + O(\eta^2),
\end{align}
where we write only the term with one derivative acting on $\xi$ and
one derivative acting on $\Sigma$.  This term, if nonzero, is the only
one that can yield a $ZH\eta$ coupling.

\begin{figure}
\centerline{\includegraphics[width=0.80\columnwidth,height=.7\columnwidth]{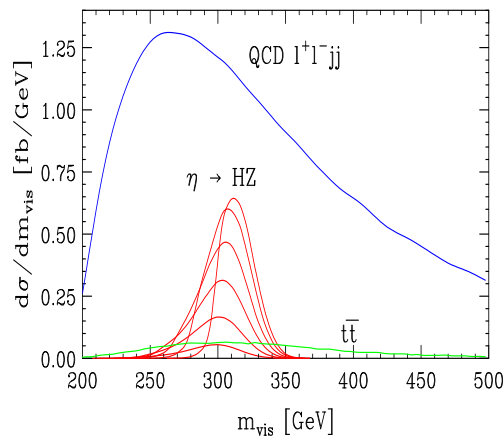}}  
\caption{Case: heavy $\eta$, light Higgs, invariant mass for the
  $\ell\ell bb$ final state for $\eta \to HZ$, $Z\to \ell\ell$ and
  $\eta\to bb$. The QCD background is blue, the top background
  green. The pseudoaxion signal is red for a mass varying from 280 to
  335 GeV.}\label{fig:mtotal}   
\end{figure} 

We now use the special structure of the covariant derivatives in
product group models, which is the key to the Little Higgs mechanism:
$D_\mu \Sigma = \partial_\mu \Sigma + A_{1,\mu}^a \left( T_1^a \Sigma 
+ \Sigma (T_1^a)^T \right) + A_{2,\mu}^a \left( T_2^a \Sigma
+ \Sigma (T_2^a)^T \right)$, 
where $T_i^a, i=1,2$ are the generators of the two independent $SU(2)$
groups, and $A_{i,\mu}^a = W^a_\mu$ + heavy fields. Neglecting the heavy
gauge fields and extracting the electroweak gauge bosons, we have
$\text{Tr}\left[ (D^\mu \Sigma)^\dagger \Sigma \right] \sim W_\mu^a
\text{Tr} \left[ (T_1^a + T_2^a) + (T_1^a + T_2^a)^* \right] = 0$. 
This vanishes due to the zero trace of $SU(2)$ generators.  The same
is true when we include additional $U(1)$ gauge group generators such
as hypercharge, since their embedding in the global simple group
forces them to be traceless as well.  We conclude that the coefficient
of the $ZH\eta$ coupling vanishes to all orders in the $1/F$
expansion.

\newcommand{\sla}[1]{/\!\!\!#1}

\begin{figure}
\centerline{\includegraphics[width=0.8\columnwidth,height=.7\columnwidth]{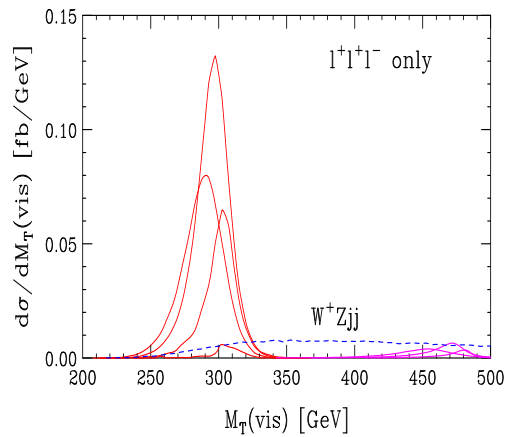}} 
\caption{Case: heavy $\eta$, heavy Higgs, process: $gg \to \eta \to
  ZH$, $Z\to\ell\ell$, $\eta\to WW$; invariant visible mass
  distribution for $\ell\ell\ell j j \sla{p}_T$. There is only a tiny
  background from $WZjj$ (dashed line).}\label{fig:mvis}  
\end{figure} 

Next, we consider the simple group models, where we
use the following notation for the nonlinear sigma fields:
$\Phi\zeta$, where $\Phi=\exp[i\Sigma/F]$ and $\zeta=\left(0,\ldots
0,F\right)^T$ is the vev directing in the $N$ direction for an $SU(N)$
simple gauge group extension of the weak group.  Thus, in simple group
models the result is the $N,N$ component of a matrix:
\begin{align}\label{kin:simplegroup}
\mathcal{L}_{\text{kin.}} &\sim\; F^2 D^\mu (\zeta^\dagger
 \Phi^\dagger) D_\mu 
 (\Phi\zeta) \nonumber\\ &=\; \ldots +  i F (\partial_\mu \eta) \left(
 \Phi^\dagger (D_\mu \Phi) - (D_\mu \Phi^\dagger) \Phi \right)_{N,N} \; .
\end{align}
We separate the last row and column in the matrix representations of
the Goldstone fields $\Sigma$ and gauge boson fields $\mathbf{V}_\mu$:
the Higgs boson in simple group models sits in the off-diagonal
entries of $\Sigma$, while the electroweak gauge bosons reside in the
upper left corner of $\mathbf{V}_\mu$. With the
Baker-Campbell-Hausdorff identity, one gets for the term in
parentheses in Eq.~(\ref{kin:simplegroup}): 
\begin{align}\label{eq:vexpansion}
  & \; \mathbf{V}_\mu + \frac{i}{F} \lbrack \Sigma, \mathbf{V}_\mu
  \rbrack - \frac{1}{2F^2} \lbrack \Sigma , \lbrack \Sigma,
  \mathbf{V}_\mu \rbrack \rbrack + \ldots \nonumber\\
  =&\; \begin{pmatrix}
    \mathbf{W}_\mu & 0 \\ 
    0 & 0 
  \end{pmatrix}
  + \frac{i}{F} 
  \begin{pmatrix}
    0 & - \mathbf{W}_\mu h \\ h^\dagger \mathbf{W}_\mu & 0 
  \end{pmatrix} \\ &\; - \frac{1}{2F^2} 
  \begin{pmatrix}
    h h^\dagger \mathbf{W} + \mathbf{W} h h^\dagger & 0
    \\ 0 & - 2 h^\dagger \mathbf{W} h 
  \end{pmatrix} + \ldots \nonumber 
\end{align}
The $N,N$ entry can only be nonzero from the third term on. The first
term, would be a mixing between the $\eta$ and the Goldstone boson(s)
for the $Z'$ state(s) and cancels with the help of the many-multiplet
structure. If the $N,N$ component of the second term were nonzero, it
would induce a $ZH\eta$ coupling without insertion of a factor $v$.
This is forbidden by electroweak symmetry.  To see this, it is
important to note that in simple group models the embedding of the
Standard Model gauge group always works in such a way that hypercharge
is a linear combination of the $T_{N^2-1}$ and $U(1)$ generators.
This has the effect of canceling the $\gamma$ and $Z$ from the
diagonal elements beyond the first two positions, and preventing the
diagonal part of $\mathbf{W}_\mu$ from being proportional to $\tau^3$.
The third term in the expansion yields a contribution to the $ZH\eta$
coupling, $(\partial^\mu\eta)h^\dagger \mathbf{W}_\mu h \sim v H 
Z_\mu\partial^\mu\eta$.

\begin{figure}
  \centerline{\includegraphics[width=0.8\columnwidth,
      height=.6\columnwidth]{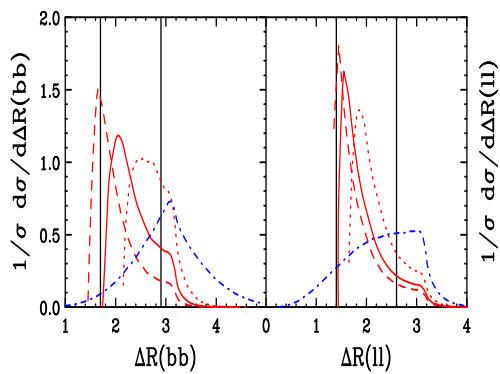}} 
\caption{Lego plot separations for $b$--$b$ and $\ell$--$\ell$ for the
  $bb\ell\ell$ final state. The signal is red (with $m_\eta$ varying
  from around 300 to 330 GeV) and the QCD background is
  blue.}\label{fig:lego}   
\end{figure} 

The crucial observation is that the
matrix representation embedding of the two non-Abelian $SU(2)$ gauge
groups, and especially of the two $U(1)$ factors within the
irreducible multiplet of the PNGBs of one simple
group (e.g. $SU(5)$ in the Littlest Higgs), is responsible for the
non-existence of this coupling in product group models.  It is exactly
the mechanism which cancels the quadratic one-loop divergences
between the electroweak and heavy $SU(2)$ gauge bosons which cancels
this coupling.  In simple group models the Higgs mass term
cancellation is taken over by enlarging $SU(2)$ to $SU(N)$, and the
enlarged non-Abelian rank structure cancels the quadratic divergences
in the gauge sector -- but no longer forbids the $ZH\eta$
coupling. Hence, its serves as a discriminator
between the classes of models.  


\section{LHC and ILC phenomenology}

The pseudoaxion(s) can be produced at the LHC in gluon fusion and 
discovered in the rare decay mode
$\gamma\gamma$~\cite{pseudoaxions}. But the $\eta HZ$ coupling can be
observed at the LHC only if either one of the decays $H \to 
Z \eta$ or $\eta \to ZH$ is kinematically allowed. This leaves large
holes in parameter space, which can be covered by a $500-1000$ GeV
ILC, depending on the masses (see below). Here, we focus on the discovery
potential of the LHC for the pseudoaxions, assuming the presence of
the $ZH\eta$ coupling. We assume the Simplest Little Higgs with
parameters chosen to fulfill the low-energy constraints. The two cases
a) $gg \to \eta \to HZ$ and b) $gg \to H \to \eta Z$ lead to similar final
states, depending on the masses of the Higgs and pseudoaxion. For
light Higgs or light pseudoaxion, a) and b) lead to the final state
$bb\ell\ell$, while case a) for heavier Higgs leads to a $ZWW \to
\ell\ell \ell jj\sla{p}_T$ final state. In the first case, there is
severe background from continuum QCD $\ell\ell jj$ production, while
the top background is manageable. We apply the following
cuts: $p_T(b) > 25~{\rm GeV}$, $|\eta(b)| < 2.5$, $p_T(\ell) > 15~{\rm
  GeV}$, $|\eta(\ell)| < 2.5$, $\triangle R(bb,b\ell) > 0.4$,
$\triangle R(\ell\ell) > 0.2$; furthermore $89.6 < m_{\ell\ell} <
92.8~{\rm GeV}$ and  \; $\sla{p}_T < 30~{\rm GeV}$ to reduce the top
background. The result for the total transverse invariant mass is
shown in Fig.~\ref{fig:mtotal}, where the Simplest Little Higgs is
shown for the Golden Point~\cite{simplest,pseudoaxions}. 
Fig.~\ref{fig:mvis} shows the total visible invariant mass for the
final state $\ell\ell\ell jj \sla{p}_T$ which covers the case of a heavy
pseudoaxion decaying to a leptonically decaying $Z$ and a heavy
Higgs. The latter decays to $WW$ with one hadronic and one leptonic
decay. For this process, the main background comes from $WZjj$ which
is not severe for the Golden Point of the Simplest Little Higgs. 

Relaxing the parameter values for the Golden Point (which gives
near-to-maximal rates but is still consistent with electroweak
precision observables~\cite{simplest,pseudoaxions}) reduces the signal
to the size of the background. Compared to other new physics scenarios
this is still a quite comfortable situation. Fig.~\ref{fig:lego} shows
the method of lego plots as a further means to discriminate between
signal and background. On the left, there is the $\Delta R$ for the
two $b$ jets, on the right for the two leptons. The shapes of these
distributions are different between signal and background and allow
for a further optimization of the cut analysis to improve the
signal-to-background ratio. However, this goes beyond the scope of
this study here.  

\begin{figure}
\centerline{\includegraphics[width=0.80\columnwidth,
  height=.6\columnwidth]{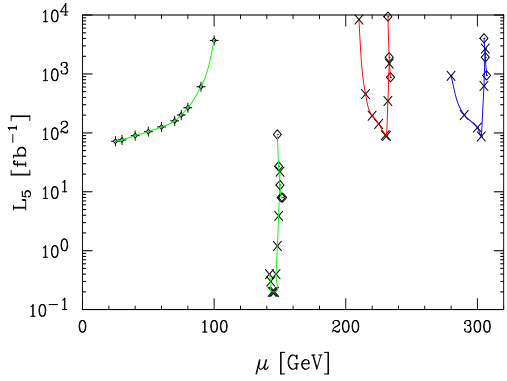}} 
\caption{Needed total integrated luminosity for the LHC to yield a
  $5\,\sigma$ $\eta$ discovery signal. Hatches are for $gg \to \eta \to
  HZ$, crosses for $gg \to H \to \eta Z$. Different colors are for
  different choices of parameters in the Simplest Little Higgs. 
  }\label{fig:massreach}  
\end{figure} 
 
Finally, Fig.~\ref{fig:massreach} shows the LHC integrated luminosity
(per experiment) needed for a 5 $\sigma$ discovery of the pseudoaxion
in Simplest Little Higgs Model using the $\eta H Z$ coupling (for
other discovery methods, cf.~\cite{pseudoaxions}). Hatches in the plot
are for $gg \to \eta \to HZ$, crosses for $gg \to H \to \eta
Z$. Different colors are for different choices of parameters in the
Simplest Little Higgs; for more details see
\cite{distinguish}. Remember, that this only holds if either of the
two decays $\eta \to HZ$ or $H \to \eta Z$ is on-shell.    

\begin{figure}
\centerline{\includegraphics[width=0.8\columnwidth,height=.6\columnwidth]
  {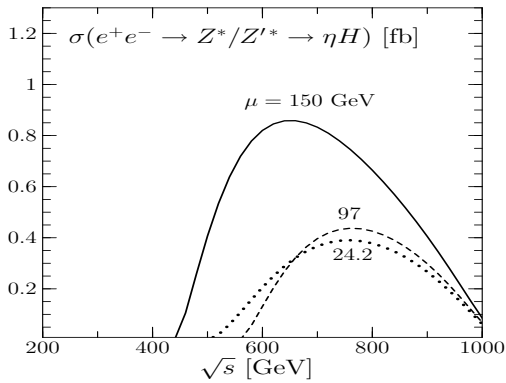}}  
\caption{ILC cross section for the $H\eta$ associated production as
  function of $\sqrt{s}$, taking into account the destructive $Z/Z'$
  interference. The full, dashed and dotted lines correspond to
  $m_\eta = 309/200/50$ GeV, respectively.}\label{fig:xsec}  
\end{figure}

At a high-energy ILC, the production happens in association with a
Higgs boson like in a two-Higgs-model. Fig.~\ref{fig:xsec} shows the
cross section as a function of $\sqrt{s}$ for three different values
of the $\eta$ mass. The simulations for these processes have been
performed with the WHIZARD package~\cite{omega,whizard,omwhiz}, which
is ideally suited for physics beyond the SM~\cite{omwhiz_bsm}. SM
backgrounds are nowhere an issue. Interesting is the $ZHH$ final state  
which is important for measuring the triple Higgs
coupling~\cite{triplehiggs}. In the SM the cross section is at the
borderline of detectability, but the rates
are larger by factors two to six in the Simplest Little Higgs with the
intermediate pseudoaxion. For more details see~\cite{distinguish,jr_lcws07}.

In conclusion, the LHC provides an ideal
environment for discovering pseudoaxions and measuring their
properties. The $ZH\eta$ coupling can be used as a tool for the
discrimination between simple and product group models. Holes in
parameter space left over by LHC can be closed by a 1 TeV ILC. 
 
\section{Acknowledgments} 
 
JR was partially supported by the Helmholtz-Gemein\-schaft under
Grant No. VH-NG-005 and the Bundes\-ministerium f\"ur Bildung und
Forschung, Germany, under Grant No. 05HA6VFB.

%
%

\end{document}